\documentclass[conference]{IEEEtran}
\IEEEoverridecommandlockouts
\usepackage{cite}
\usepackage{amsmath,amssymb,amsfonts}
\usepackage{algorithmic}
\usepackage{graphicx}
\usepackage{textcomp}
\usepackage{xcolor}
\def\BibTeX{{\rm B\kern-.05em{\sc i\kern-.025em b}\kern-.08em
    T\kern-.1667em\lower.7ex\hbox{E}\kern-.125emX}}
\begin{document}

\title{Workflow-Driven Distributed Machine Learning in CHASE-CI: A Cognitive Hardware and Software Ecosystem Community Infrastructure \\
\thanks{Corresponding author: Ilkay Altintas, email: ialtintas@ucsd.edu. This work was supported by NSF-1730158 for Cognitive Hardware and Software Ecosystem Community Infrastructure (CHASE-CI), NSF-1541349 for The Pacific Research Platform (PRP), and NSF-1826967 for The National Research Platform (NRP).}
}

\author{
\IEEEauthorblockN{
Ilkay Altintas\IEEEauthorrefmark{1},
Kyle Marcus\IEEEauthorrefmark{1},
Isaac Nealey\IEEEauthorrefmark{2},
Scott L. Sellars\IEEEauthorrefmark{2},
John Graham\IEEEauthorrefmark{2},\\
Dima Mishin\IEEEauthorrefmark{1},
Joel Polizzi\IEEEauthorrefmark{2},
Daniel Crawl\IEEEauthorrefmark{1},
Thomas DeFanti\IEEEauthorrefmark{1},
Larry Smarr\IEEEauthorrefmark{2}
\IEEEauthorblockA{
\IEEEauthorrefmark{1}\textit{San Diego Supercomputer Center},
\textit{University of California San Diego},
La Jolla, CA, USA\\
\IEEEauthorrefmark{2}\textit{Qualcomm Institute},
\textit{University of California San Diego},
La Jolla, CA, USA
}
}
}


\maketitle

\begin{abstract}
The advances in data, computing and networking over the last two decades led to a shift in many application domains that includes machine learning on big data as a part of the scientific process, requiring new capabilities for integrated and distributed hardware and software infrastructure. This paper contributes a workflow-driven approach for dynamic data-driven application development on top of a new kind of networked Cyberinfrastructure called CHASE-CI. In particular, we present: 1)  The architecture for CHASE-CI, a network of distributed fast  GPU  appliances  for  machine  learning  and  storage managed through Kubernetes on the high-speed (10-100Gbps) Pacific Research Platform (PRP); 2)  A machine learning software containerization approach and libraries required for turning such a network into a distributed computer for big data analysis; 3) An atmospheric science case study that can only be made scalable with an infrastructure like CHASE-CI; 4) Capabilities for virtual cluster management for data communication and analysis in a dynamically scalable fashion, and visualization across the network in specialized visualization facilities in near real-time; and, 5) A step-by-step workflow and performance measurement approach that enables taking advantage of the dynamic architecture of the CHASE-CI network and container management infrastructure.
\end{abstract}


\section{Introduction}

Over the last two decades, massive changes occurred in data collection and analysis. These new advances including on demand computing, Big Data and the Internet of Things, and new forms of machine learning, lead to a new age for Artificial Intelligence with a high dependency on networking and connectivity. All fields of scientific research have also observed major changes in how science is being conducted today, creating requirements for cyberinfrastructure (CI) that is as dynamic and data-driven as the science it supports. Scientific endeavor today often includes distributed scientific applications running on a continuum of computational resources with the need for near real-time big data processing capabilities to process data streaming from remote instruments or large scale simulations. Due to such influences, one of the most rapidly growing core research fields is machine learning (ML) with large datasets, either static or streaming. 


Research at the frontier of this emerging discipline requires use of large amounts of compute time (more and more on Graphics Processing Units, GPUs) and specialized non-von Neumann (NvN) processors, along with ability to use, measure and scale a rapidly growing number of software libraries and technologies while moving data from its source rapidly for archival and processing. 

Moreover, as the number of applications that need such capabilities grew, a need emerged for development tools and user-facing environments to interact with these new forms of CI and ML, and build scientific applications on top of them. A big challenge here is the integration of CI capabilities (including a networks stack, ranging from high-level tools and interfaces, down to low-level hardware) in a way that matches the dynamic needs of executing applications in user workflows. 

The United States National Science Foundation (NSF) funded two data movement and storage CI projects, the Pacific Research Platform (\textit{PRP}) and its successor National Research Platform (\textit{NRP}), to build a new research CI that addresses all of the following aforementioned challenges and needs: 
\begin{enumerate}
\item The ability to share affordable GPU resources among many researchers; 
\item Exploitation of new-generation energy efficient NvN processors;
\item Access to a wide array of ML algorithms; 
\item Access to, or the ability to rapidly gain, expertise in managing and measuring such systems; and, critically, 
\item The facilitation of rapid access, movement and storage of extremely large datasets.
\end{enumerate}

A third NSF project called \textit{CHASE-CI}, a Cognitive Hardware and Software Ecosystem Community Infrastructure, was funded to enable deployment, measurement and utilization of machine learning libraries and storage on top of the PRP infrastructure, and  development of user workflows that can dynamically be measured and configured on top of the deployed network, hardware and software. 

\textbf{Contributions.} The discussion in the paper relates to this need for and importance of facilitating user workflows on top of the existing and up and coming dynamic 
networks for distributed advanced computing. In particular, we present:  
\begin{enumerate}
\item The architecture for CHASE-CI, a network of distributed fast GPU appliances for machine learning and storage managed through Kubernetes on the high-speed (10-100 Gbps) Pacific Research Platform (PRP),
\item A machine learning software containerization approach and libraries required for turning such a network into a distributed computer for big data analysis,
\item An atmospheric science case study that can only be made scalable with an infrastructure like CHASE-CI, 
\item Capabilities for virtual cluster management for data communication and analysis in a dynamically scalable fashion, and visualization across the network in specialized visualization facilities in near real-time, and 
\item A step-by-step workflow and performance measurement approach that enables taking advantage of the dynamic architecture of the CHASE-CI network and container management infrastructure.
\end{enumerate}

\textbf{Outline.} The rest of this paper is organized as follows. In Section~\ref{sec:chaseci}, we introduce the CHASE-CI infrastructure and explain our container management approach in CHASE-CI (contributions 1 and 2). Section~\ref{sec:usecase} introduces the atmospheric science study and describes each step of a machine learning workflow within this study in detail (contribution 3). In Sections~\ref{sec:ns}~and~\ref{sec:pod}, we introduce the Kubernetes constructs for virtual cluster and resource management using namespaces and pods (contribution 4).We introduce a collaborative workflow integration, measurement and execution approach (contribution 5) in Section~\ref{sec:wf} . We review our other related work in Section~\ref{sec:rw} and conclude in Section~\ref{sec:conc}.

\section{CHASE-CI Infrastructure} \label{sec:chaseci}
The CHASE-CI project takes advantage of existing CI to put machine learning tools at the fingertips of researchers. High-bandwidth data transfer and access to GPUs provides a framework for a variety of machine learning workflows. The hardware backbone of CHASE-CI is the Pacific Research Platform (PRP), a partnership of more than 20 institutions, including four NSF/DOE/NASA supercomputer centers \cite{c1}. 
By deploying Data Transfer Nodes (DTNs) at partner sites, the PRP (with the support of CENIC \cite{c2}) established a high-speed cloud connected on 10G, 40G and 100G networks using the ESnet Science DMZ \cite{c3} model as a basis for its architecture. The Science DMZ model consists of simple, scalable networks with a focus on security and high-performance computing. DTNs are responsible for the efficient movement of large amounts of scientific data to and from remote sites. Performance is optimized by minimizing data transfer on Local Area Networks (LANs), favoring high-bandwidth Wide Area Networks (WANs) such as the fiber connecting PRP sites. 

\begin{figure}[h]
\centering
\includegraphics[width=0.5\textwidth]{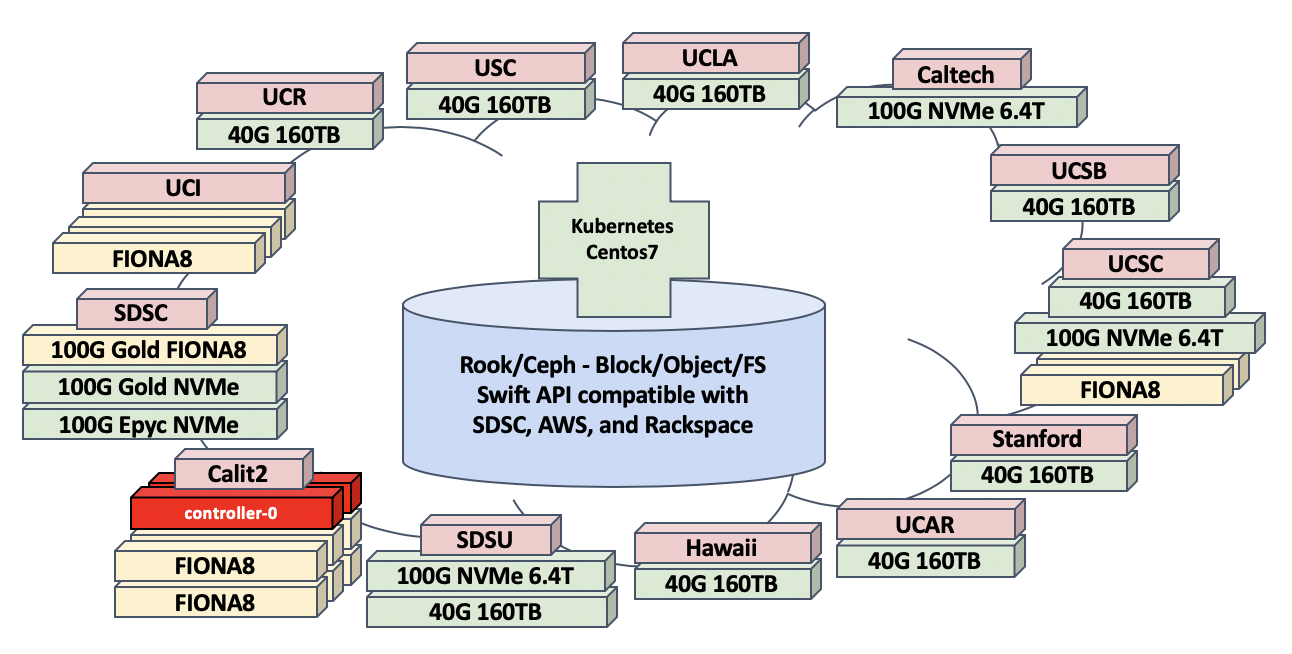}
\caption{Running Kubernetes/Rook/Ceph On PRP Allows the Deployment of a Distributed PB+ of Storage for Posting Science Data}
\label{fig:fionas}
\end{figure}

Data Transfer Nodes at PRP endpoints are named Flash I/O Network Appliances (\textit{FIONAs}) and, as the name implies, contain high performance Network Interface Controllers (NICs) and high capacity Solid State Drives (SSDs). They can be built to different specifications depending on budget and network capabilities. The basic FIONA machines at Calit2 each contain dual 12-core CPUs, 96 GB RAM, 1TB SSD, and two 10 GbE interfaces \cite{c1}. CHASE-CI adds clouds of game GPUs and NvN machines to the PRP. Multi-tenant, ``FIONA8'' machines containing eight game GPUs each have been installed at various PRP sites, along with over a petabyte of storage (SSD and NVMe) for hosting scientific data. Using a containerized, self-healing ecosystem built with open source tools laid the framework for promoting workflows in an extensive hyperconverged system named ``Nautilus''. Figure~\ref{fig:fionas} shows the FIONA8 and storage nodes distributed on the PRP backbone and managed dynamically through containerization. 

\subsection{Container Orchestration in CHASE-CI}
Kubernetes is the container orchestration engine used for management and deployment in Nautilus. It is an increasingly popular tool first open-sourced by Google in 2014 \cite{c4}. Using containers guarantees environmental consistency, resource isolation, and portability. The same container can run on a variety of systems, as each containerized application deploys with its software dependencies. A Kubernetes cluster consists of hardware resources (compute and memory), containerized applications, and a set of policies which define the desired behavior of the cluster. The cluster is further split into different namespaces that allow for separation of deployment execution. This namespace is a grouping of containers that can all talk to each other and can share a common data store. A framework is provided to define policies pertaining to networking, security, load balancing, fault tolerance, updates, and resource management. In addition to policies for scheduling containers, software is needed to run the containers. The Nautilus cluster uses the open-source Docker platform for operating-system-level virtualization, i.e. ``containerization'' \cite{c5}. 

As many devices are not supported natively in Kubernetes, a device plugin is deployed to provide low-level access to CHASE-CI GPUs from within a container \cite{c6}. Using these runtime hooks, researchers can leverage the hardware using the CUDA API \cite{c7} to drastically reduce the amount of compute time needed for data analytics and training artificial intelligence algorithms on large data sets. Many machine learning projects use a Jupyter Notebook to run their hardware accelerated training using high-level tools such as Tensorflow and PyTorch, and these are easily deployed on multi-GPU nodes. However, using custom containers, such as the case study discussed in Section~\ref{sec:usecase}, allows powerful scalability and granular control of the workflow.

Vast amounts of fast storage is paramount to the efficiency of machine learning workflows. As shown in Figure~\ref{fig:fionas}, Nautilus uses Rook, an embedded strain of the Ceph cloud-native storage system \cite{c8}. Ceph provides block, object, and POSIX compliant file storage as a service within the cluster. Massively scalable, Ceph replicates and dynamically distributes data between storage nodes while monitoring their health. Based on RADOS \cite{c9}, Ceph is largely autonomous and ensures high availability. Data is easily mounted and shared between containers running in Nautilus, as well as compatible with other cloud storage solutions such as Amazon S3, OpenStack Swift, and various supercomputer storage architectures via the Ceph Object Store \cite{c10}, e.g., at the San Diego Supercomputer Center (SDSC).

Equipped with a cloud of compute and storage nodes, hundreds of GPUs, and kilometers of high-bandwidth fiber paths between institutions, Nautilus needs software to monitor the health, availability, and performance of resources. Grafana is an open source platform for time series analytics \cite{c11}. It graphs cluster health and performance data using a functional query language provided by Prometheus \cite{c12}. Grafana’s web-based dashboard (e.g., Figure~\ref{fig:fig1}) is accessible from a browser, providing a quick debugging solution for cluster users and administrators.

\section{A Chase-ci Case Study: Object Segmentation Workflow}\label{sec:usecase}
Often, earth science phenomena (e.g., rain clouds, flash floods, droughts, wildfires, ocean temperatures) are not clearly defined and change dynamically in time and space, making it challenging to apply rapid object segmentation to the earth sciences. Most segmentation approaches, including Deep Learning algorithms, only extract pixel-level segmentation masks, and typically do not consider the temporal information of the data. However, the CONNected objeECT, or CONNECT algorithm \cite{c20, c21} 
focuses on keeping track of the entire life-cycle of a detected earth science phenomena by “connecting” pixels in time and space. Previous work on the CONNECT algorithm focused on using MATLAB functions using a single CPU, limited memory, and storage hardware, and no access to Data Transfer Nodes (DTNs) on a high speed research network. To improve on this approach, we used CHASE-CI resources to accelerate the CONNECT workflow. We experimented with Machine Learning approaches that were optimized for GPU acceleration in order to do rapid object segmentation using National Aeronautics and Space Administration (NASA) data. 

The CHASE-CI resources provided unique capabilities with the combination of a high speed research network, FIONA8s, Ceph Object Store, and accessible GPUs using Kubernetes orchestration. These capabilities allow for flexible workflow environments, at-scale machine learning for object segmentation, and automated deployment across the CHASE-CI kubernetes cluster. The combination of these resources innovated the CONNECT workflow in several ways, including: 
\begin{enumerate}
\item \textit{Rapid data transfer between a FIONA and Ceph cloud-based object storage, which is distributed across the Pacific Research Platform (PRP):} This is performed harnessing  Unidata's Thematic Real-time Environmental Distributed Data Services (THREDDS) \cite{c19} server maintained on a node within the PRP allowing Kubernetes to transfer data into the Nautilus system. 
\item \textit{Applying a new object segmentation algorithm: }Instead of using MATLAB functions that use a single CPU to do the object segmentation, a new algorithm, Flood-Filling Network (FFN) \cite{j17}, was used. The FFN was applied to NASA data using 50 NVIDIA 1080ti GPUs based on Tensorflow. For this case study, 455GB of 3-hourly, NASA Modern-Era Retrospective Analysis for Research and Applications, Version 2 (MERRA V2) dataset from January 1, 1980 to May 31, 2018 was used. The MERRA V2 reanalysis product represents the satellite era using a state-of-the-art assimilation system, known as the Goddard Earth Observing System Data Assimilation System Version 5 \cite{c18, c16}. The data has a temporal frequency of 3-hourly from 00:00 UTC (instantaneous), with a 3-D spatial grid at full horizontal resolution. The resolution is 0.5° x 0.625° in latitude and longitude (i.e., global resolution of 576x361 pixels), and 42 vertical levels in the atmosphere. This data is used in this case study for calculating Integrated Water Vapor Transport (IVT) from the assimilated meteorological field data archive (M2I3NPASM).
\end{enumerate}

\begin{figure}[h]
\centering
\includegraphics[width=.35\textwidth]{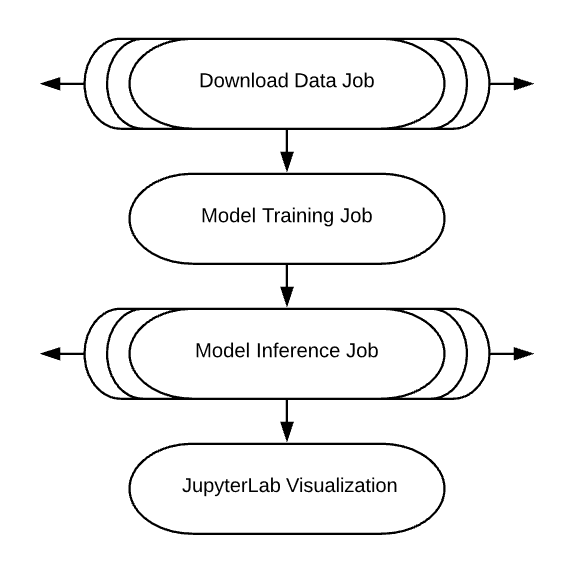}
\caption{Workflow steps}
\label{fig:fig4}
\end{figure}

The accelerated workflow was developed to use multiple Docker images for job specific tasks. As illustrated in Figure~\ref{fig:fig4}, the steps taken in the accelerated workflow include: 1. downloading data from THREDDS and data preparation, 2. model training, and 3. distributed multi-GPU model inference. Step 4, the final step, is visualization. In addition, the Nautilus Grafana dashboard was used to monitor jobs at each step of the workflow and is reported below. Over the next four subsections, we describe a step-by-step description of the accelerated workflow.

\subsection{Step 1: THREDDS Data Download}
Three Docker images were created for handling the downloading steps necessary for transferring data to the Ceph Object Store (which can be seen by all nodes in the custer). THREDDS is a web server that provides metadata and data access for scientific datasets using a variety of remote data access protocols \footnote{THREDDS remote data access protocol catalog: \\ http://its-dtn-02.prism.optiputer.net:8080/thredds/catalog.html}. THREDDS provides a data subset tool that allows for selection of a variable within files, and provides the capability to transfer only that specific variable instead of the entire file. Using this capability (for this case study we selected \textit{Integrated Water Vapor Transport (IVT)}) we reduced our total archive size from 455GB to 246GB. This allowed us to significantly reduce the need to download entire files with many variables and metadata, greatly increasing the speed at which data is transferred. 

\begin{figure}[h]
\centering
\includegraphics[width=.5\textwidth]{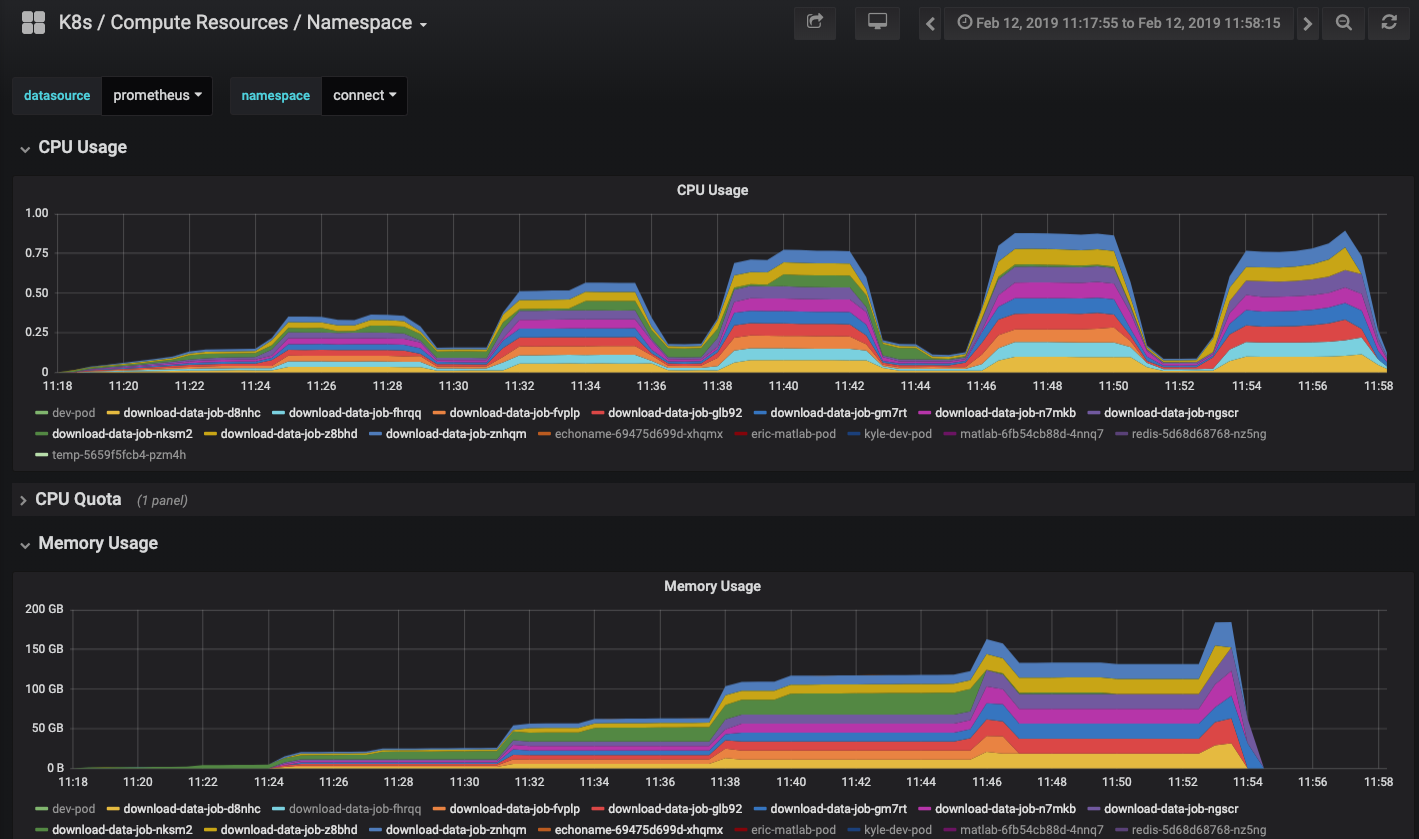}
\caption{Kubernetes data download job orchestration: 10 Workers, managed by a Redis job queue (each color represents a worker). Total time to run is 37 minutes with a total data size transfer of 246GB (112,249 NetCDF files). Graph shows CPU and Memory usage during this time.}
\label{fig:fig1}
\end{figure}

\begin{figure}[h]
\centering
\includegraphics[width=0.5\textwidth]{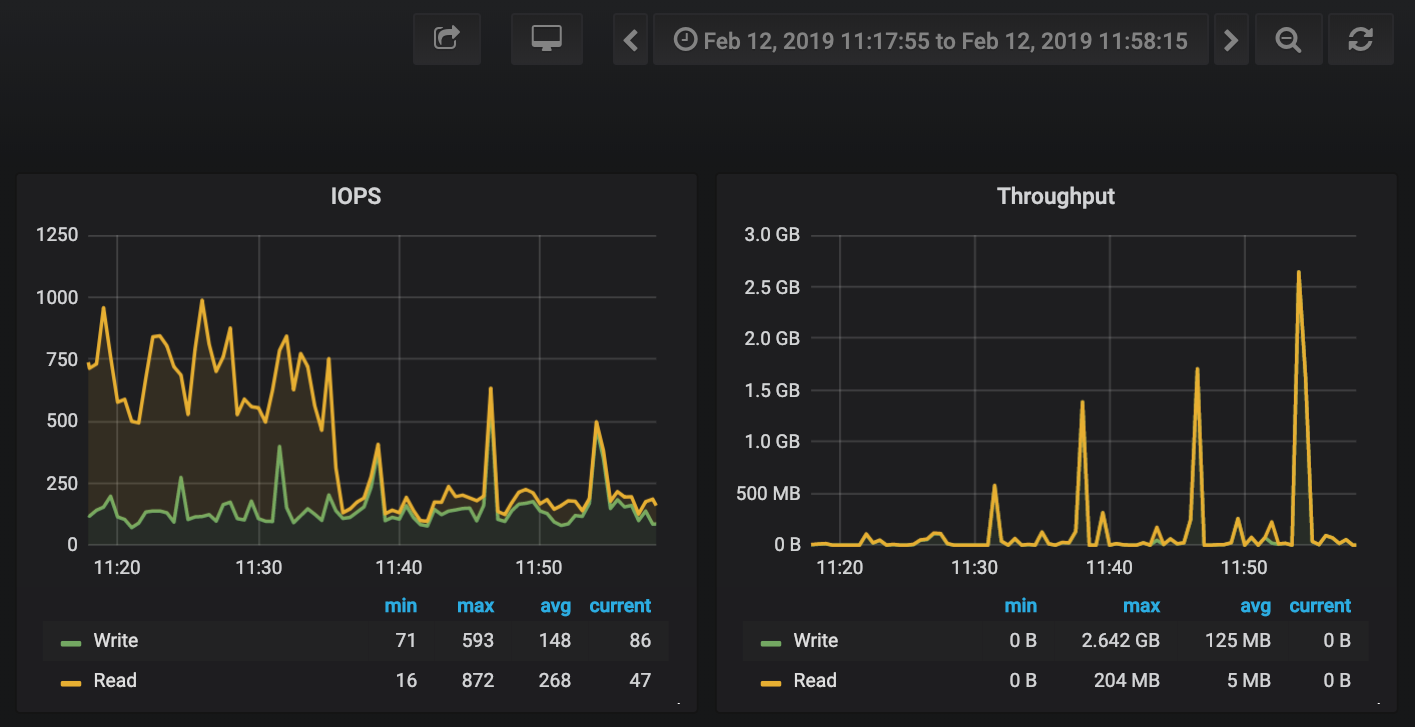}
\caption{Network usage during download job run. IOPS: Max 593MB/s. Throughput: Max 2.64GB}
\label{fig:fig2}
\end{figure}

The first step uses 10 Kubernetes workers (pods) and were set up through a Job resource to download the 246GB (112,249 NetCDF files) from THREDDS. The Kubernetes Job handles how each worker pod is set up and allows for easily scaling the number of workers present. The Job also handles creating pods on different nodes if pods are shut down by the system or crash.  To execute the data download, each worker uses the open source Aria2\footnote{Aria file transfer software: https://aria2.github.io/} file transfer software that allows multiple parallel downloads (20 parallel downloads in our case) to retrieve urls stored in a list of data files streamed from a Redis queue. The Redis queue holds a list of files that contain urls to download from the THREDDS server, each pod pops a message off the queue and uses the file path included in the message as input to Aria2. Aria2 then downloads all the urls listed in the file.  

The Redis queue was developed to keep track of which files were downloaded and to distribute the work across pods. The workers continue to process messages in the Redis queue until all files within the Redis database have been downloaded. With 112,249 NetCDF files in total, each worker also merges the small individual files into larger (Hierarchical Data Format) files for input into the FFN model and transfers the larger file to the Ceph Object Store. Once completed, the data has been transferred from an online archive, processed, and stored in the Ceph Object Store for future Kubernetes training and inference jobs. 

Step 1's total run time is 37 minutes with IOPS of 593MB/s (max) and throughput of 2.64GB (max) as seen in Figures~\ref{fig:fig1} and \ref{fig:fig2}. Each color in Figure~\ref{fig:fig1} represents an individual worker.

\subsection{Step 2: Model Training}
Once the data has been transferred to the storage volume (CephFS accessible by all nodes) and the data has been split up into the appropriate number of subsets, a single additional Docker image is spawned for model training. The model selected to do rapid segmentation was the FFN model and was adapted to do segmentation of NASA data.

\begin{figure}[h]
\centering
\includegraphics[width=0.48\textwidth]{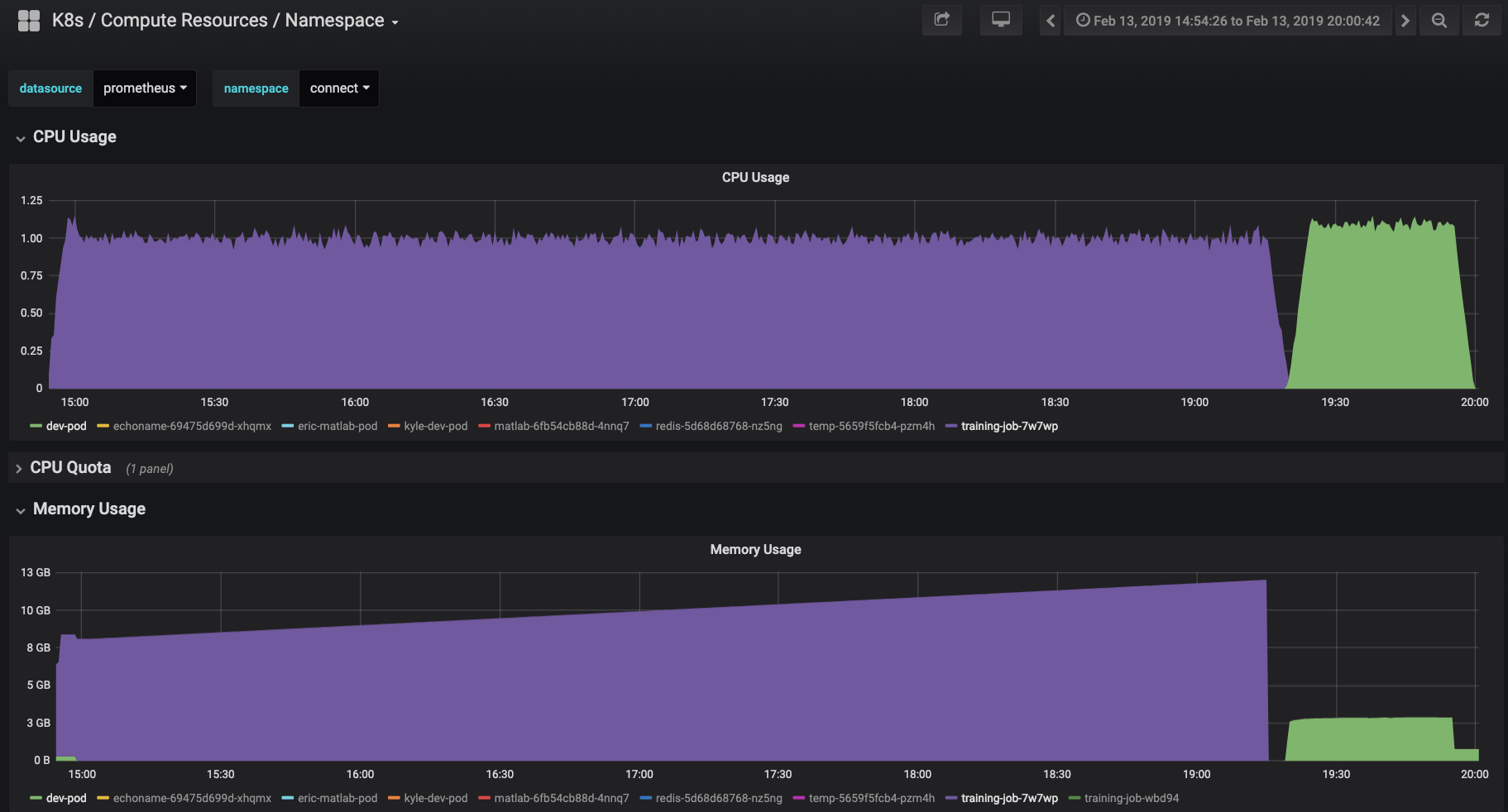}
\caption{Training job - Purple shows the data preparation job. Green is the FFN algorithm training on a 576x361x240 data volume. }
\label{fig:fig3}
\end{figure}
 
The FFN model uses a 3D convolution neural network (3D CNN) developed by Google based on Tensorflow. The 3D CNN is able to separate objects within a 3D volume of spatial data or images by using a deep stack of 3D convolutions \cite{j17}. The network is trained to take an input object mask within the network’s field of view to infer the boundaries of the objects. Training the model relies on a labeled dataset, differentiating the objects with numerical values representing categories of objects. In our case, it is a binary representation of locations on earth where intense large-scale moisture transport (IVT) processes exist. The CONNECT dataset \cite{c22} is used for training, which includes segmented IVT objects in binary label representation. These labeled objects are then used to train the FFN for 30 days of data (240 3-hourly  images) and a file size of 381MB providing a training volume of 576x361x240 voxels. The training was performed on a single NVIDIA 1080ti GPU using CUDA 9 and Tensorflow 1.13.0-rc1. A detailed description of the model is beyond the scope of this paper and FFN model and parameters can be reviewed on github (https://github.come/ffn)\cite{j17}. Step 2's total run time is 306 minutes. Figure~\ref{fig:fig3} shows the performance comparison between data preparation and training using the FFN algorithm on a 576x361x240 data volume.

\subsection{Step 3: Model Inference}

The trained FFN model is then saved in the Ceph Object Store, including all parameters and configurations needed to do inference on new NASA data. Depending on the number of GPUs available at the time (in this example 50 NVIDIA 1080ti GPUs were used) a series of kubernetes run files are generated in order to distribute the inference job to many workers, each with a dedicated GPU and subset of the new NASA data. The number of GPUs in this section can scale to any number depending on the number of inference jobs needed. This is where it’s using the model previously generated from the training step. The entire 246GB (576x361x112,249 or 2.3e10 voxels) is evenly distributed across the 50 GPUs and the total inference time is 18 hours 53 minutes (1133 minutes). Ongoing experiments with model settings are expected to improve the inference time and will be reported on in future publications.

\begin{figure}[h]
\centering
\includegraphics[width=0.48\textwidth]{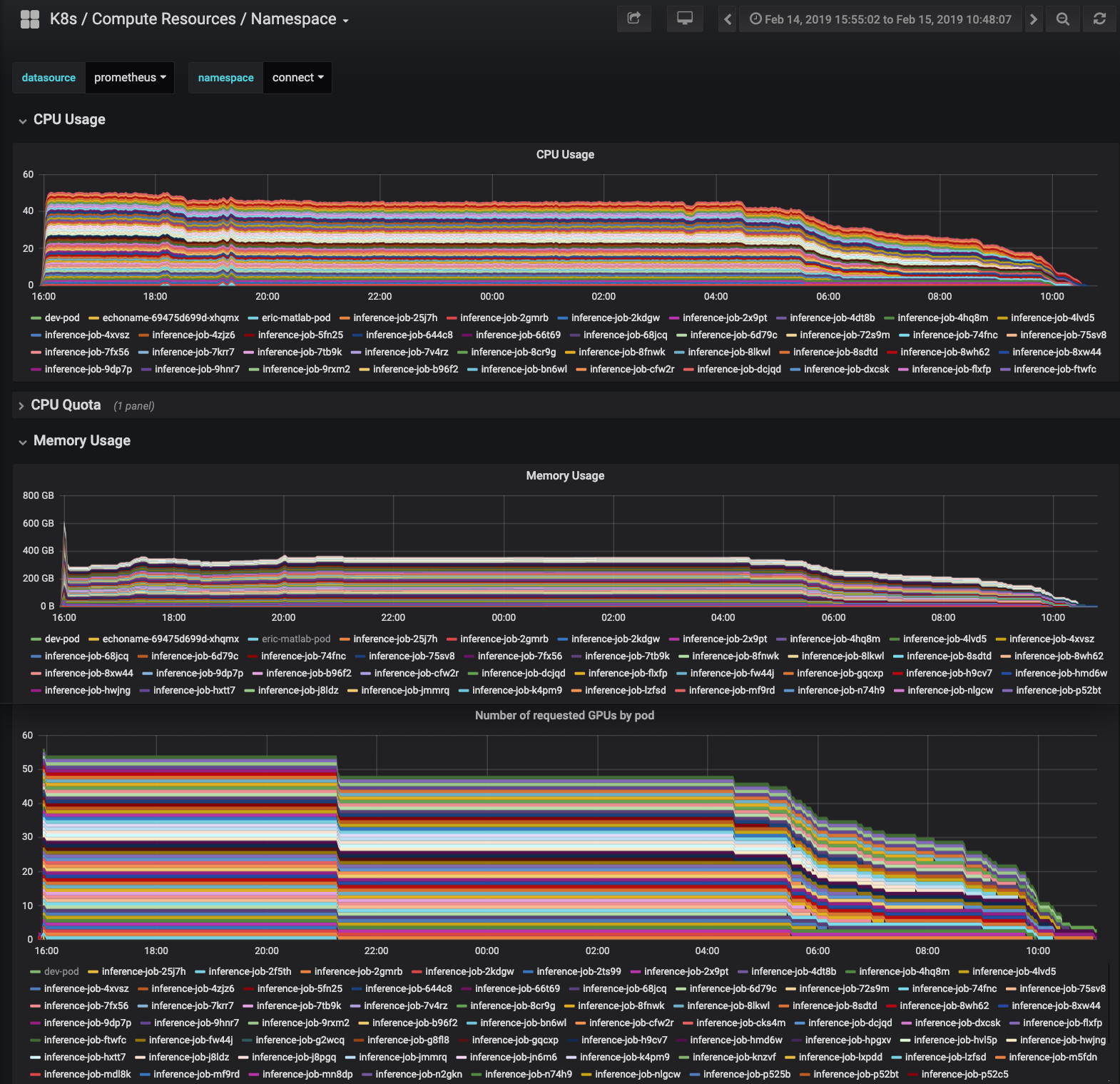}
\caption{Inference job - Top) Number of CPUs being utilized, Middle) Memory utilization, Bottom) Number of GPUs being utilized.}
\label{fig:fig4}
\end{figure}

Note that the training volume is removed from the test data volume for all validation metrics. Because Nautilus contains so many GPUs, CHASE-CI is an ideal situation and environment to run this step. It would take a long time for a limited number of GPUs to produce the same result without access to this GPU cluster.

\subsection{Step 4: JupyterLab Visualization}

The Ceph Object Store contains all workflow files and model results, which allows for efficient analysis of results and visualization using Nautilus's JupyterLab\footnote{JupyterLab Environment: https://github.com/jupyterlab/jupyterlab}. JupyterLab can be accessed with just a web browser and mitigates the need to do large scale data results file transfers to a local machine for post-processing analysis. Using a web browser, a Nautilus JupyterLab instance can be spawned with the Ceph Object Store mounted. A visualization notebook is then used to load the  most recent results, plot out the segmented objects, and calculate object statistics for post-processing analysis. The researcher benefits greatly due to the ability to quickly access the model results while also accessing the flexible resources provided by the Nautilus cluster, allowing the researcher to quickly make adjustments to the data, model, variables, and parameters and then rerun the experiment. 

\begin{table}[]
\centering
\caption{Nautilus resource summary table for all steps in the workflow}
\begin{tabular}{|l|l|l|l|l|}
\hline
               & Step 1 & Step 2 & Step 3 & Step 4 \\ \hline
\# of Pods     & 14     & 1      & 50     & 1      \\ \hline
\# of CPUs     & 42     & 1      & 50     & 1      \\ \hline
\# of GPUs     & 0      & 1      & 50     & 1      \\ \hline
Data Processed & 246GB  & 381MB  & 246GB  & 5.8GB      \\ \hline
Memory         & 225GB  & 14.8GB   & 600GB & 12GB     \\ \hline
Total Time     & 37m & 306m      & 1133m      & NA      \\ \hline
\end{tabular}
\end{table}

\subsection{Workflow Summary and Future Steps}

Each of the workflow steps described in this case study were developed to be deployed "at-scale" and adaptable to new resources and Machine Learning models as they come online within Nautilus. As described in Step 2 and shown in Figure \ref{fig:fig3}, model training is slow, including the pre-training steps of building training partition volumes and data coordinates for the FFN model. This is a challenge and is an area of active research, but once a model is trained, the inference step allows for rapid object segmentation at scales not possible without access to a cluster of GPUs like the ones that are part of Nautilus. 

We continue to optimize the execution of this workflow and add new strategies based on the measurements within CHASE-CI and other cloud environments through Kubernetes. In addition, the design of the workflow does provide the capabilities of including other Machine Learning algorithms. Here, we summarize our ongoing work towards planned workflow extensions.

\subsubsection{Distributed Data Pre-processing} 
With any machine learning process, the data pre-processing step is always a large one. TensorFlow is the backend framework used for training this model and the input to this system is translated from NetCDF files to a binary representation in a protocol buffer file (protobuf) format. This file representation is used to structure the data and quickly access it in a serialized form. Currently, this file input generation process is produced through a serial process that creates the protobuf file. However, using new advances in Tensorflow API and the Nautilus infostructure, this can be modified to distribute this work in parallel to many worker jobs. This would greatly decrease the time it takes to make these input files. To accomplish this, the input NetCDF files that need to be processed will be put into a queue and distributed to worker jobs on the cluster. These jobs will be handled by Kubernetes and will be able to scale up to any needed number of jobs very easily by just changing the scaling configuration of the Job structure. All the jobs will produce their own protobuf output and then store it in the attached CephFS directory that all nodes in the namespace can see. All of these separate protobuf files can then be read in and combined back together when setting up the training step. In summary, this allows for much faster data pre-processing and would in the future allow for training on more data since this process can scale and handle the load on the system.
\subsubsection{Distributed Training} Currently, the training on the data sets provided is being done on a single GPU node instance in Nautilus. Tensorflow does support distributed training\footnote{Distributed TensorFlow: https://www.tensorflow.org/deploy/distributed} and we want to take advantage of this. In order to accomplish this, Tensorflow and Kubernetes will have to establish a set of pods that can support this parallel training job. This would be done by first creating a Kubernetes ReplicaSet, which contains a spec to run TensorFlow distributed training clients. A ReplicaSet would be used because we would have a single client image that would need to be scaled in order to take advantage of the distributed training. Using the dynamic nature of Kubernetes, we will work on scaling it up and down depending on our needs and use the connected network to establish communication between the pods. Hostnames will be used instead of IP addresses by creating a service and providing a much more dynamic way of communicating to a pod even if its IP address changes. Once that is set up, Tensorflow will be able to distribute the training set and train in parallel. This in turn would speed up the time it takes to complete the training step and give the ability to go through the workflow faster. 
\subsubsection{Hyperparameters and Validation Datasets} 
When doing machine learning, it is important to separate training and test data. This is to avoid training on the test data for better modeling. It is also important to evaluate hyperparameters of the model. A Redis queue is being developed to store model training/testing validation split methodologies and parameters sets to be used in multi-model validation. A full object segmentation comparison is being actively worked on and is in preparation, including developing new validation data sets, looking at specific events in time and geographic regions. 

\subsubsection{Visualization} 
Since the output is a hyper volume representation, for future work we like to include the more advanced visualization part inside of the workflow as the last step. This would include using different 3D rendering packages (such as Python Mayavi 3D rendering package or Python ipvolume 3D rendering package) to display the results of the ML workflow in near real time to the user. This could even go as far as displaying the results on a large scale visualization system that runs on Nautilus, such as the SunCAVE. The visualization data could be piped through the same kubernetes orchestration software to the display on the SunCAVE and could be rendered out as data comes in.

\subsubsection{Kepler 3.0 Workflow with PPODs} 
Currently, the workflow is set up as a series of kubernetes jobs that can be controlled either through interacting with kubernetes directly or through a Jupyter Notebook that can control each step of the process. In the future we would like to move this towards a collaborative workflow using the PPODS methodology and the new Kepler 3.0 interface \cite{Kepler,Kepler2}. This would promote the collaboration effort of workflow design in a scientific community setting. It would also allow this workflow to be easily extended and tested through an educational lens.

\section{Namespace Management} \label{sec:ns}
Supporting machine learning research in multiple disciplines across several campuses is an administrative challenge. Fortunately, Kubernetes provides a framework to separate projects while providing access to the same hardware resources. A ``Namespace'' is a virtual cluster hosted within the physical cluster being orchestrated by Kubernetes \cite{c4}. Namespaces divide the cluster resources between the set of users, providing the capability to organize and segment the needs for each project into its own virtual subsection of the cluster. They provide an independent scope for names, management, and policies while scheduling jobs on the same hardware. Even though two containers may be running on the same physical machine, their affiliation to different namespaces means they are isolated from one another and may be obeying a vastly different set of resource policies or constraints. This facilitates the creation of ``user communities'' which are loosely, but not necessarily, grouped by project. Generally, the PI of a given research group is granted the role ``namespace administrator'', responsible for managing the users and resources involved in their research. Networking across namespaces is possible but requires fully qualified domain names. Low-level Kubernetes resources, such as Nodes and Persistent Volumes, are not in the scope of any namespace.

While cluster segmentation by namespace provides the chance for independent authentication policies, Nautilus uses CILogon \cite{c13} for authentication across all namespaces. CILogon is an NSF-funded, open-source authentication tool designed to federate identity across multiple authentication management systems. It provides a low barrier of entry for prospective Nautilus users, as over 2500 identity providers are supported, allowing the use of home or campus credentials. In this way, new users log on and ``claim'' their identity, rather than creating a new one. Once authenticated, an administrator can add them to their namespace from a web portal. 

\section{Nodes and PODS Management} \label{sec:pod}
Before pod orchestration frameworks such as Kubernetes existed, a lot more emphasis was put on node management when running a job or workflow.  However, Kubernetes handles a lot of the node and pod management, making the task of job management and workflow coordination an optimization. The workflow manager specifies the state configuration and passes it on to Kubernetes, and Kubernetes creates the specified state in its system.  Kubernetes will then start monitoring to make sure that the state specified is always correct and correct internal systems when needed.

For example, when running the large download job in the CONNECT workflow, the workflow manager tells Kubernetes that a certain number of worker pods is needed to download files and submit that job.  It is important to note that Kubernetes configures and manages the resources in an automatic fashion based on the specification of what is needed, i.e., it does not need a specification of "how to do it".

There are many different types of resources that can be generated in Kubernetes. For a workflow it is usually the Job resource that is most prevalent because it can execute batch process at scale.  Kubernetes will monitor these jobs which in themselves create and run pods.  The Pod is one of the most fundamental pieces of Kubernetes but it is usually not created directly. It is recommended to use Kubernetes scheduling controllers (such as Jobs or ReplicaSets) \cite{c4} because they can keep track of pods re-spawn them if any errors occur during execution.

The CHASE-CI infrastructure is very dynamic in the fact that nodes can join and leave the cluster at any time.  Kubernetes abstracts away this movement and it usually has no effect on running pods if they are set up correctly.  If a node is taken offline the pods on that node will be rescheduled on another node.

\section{Collaborative Workflow Measurement, Integration and Execution} \label{sec:wf}

Integrating and developing a workflow in a large scale environment such as the Kubernetes GPU cluster in CHASE-CI can be difficult when it's applied to a group of collaborating developers or scientists.  It's necessary to keep everyone on the same track but allow for diversified execution plans and experimentation through effective collaboration. For this purpose, we have created the PPoDS methodology to empower computational data science teams with effective collaboration tools during the exploratory workflow development phase. PPoDS stands for ``Process for the Practice of Data Science''. 

We developed a web-based CHASE-CI interface to enable the use of the PPODs methodology to transform this workflow into a interactive execution plan with the list of steps connected to each other in a visual and meaningful way, along with a set of tools for measuring and testing the development of each individual step in an analytical process towards integration. We are currently developing the tools for capturing, measuring, collecting and analyzing performance metrics during exploratory workflow development and testing process. 

The execution of the workflow needs to support the separation of steps so that each step can easily be tested independently of one another. Each step can also be developed without the concern of impeding on other workflow steps. Development can happen in parallel and brought back to execute together with the whole workflow whenever needed.

In the specific case of the CONNECT workflow, the workflow is already split up into multiple parts which allows it to scale so well. If it was instead a monolith application, it would be much harder to scale because it would first have to be refactored and split out into multiple pieces.  One of the keys that makes the CONNECT workflow so successful in scaling its ability to use worker jobs to split up individual tasks.  This allows it to easily scale in a environment that allows for dynamic resource allocation.

As a part of our future work, in order to encompass the PPoDS methodology, we would also like to add in testing to this workflow.  Creating tests for each piece of the workflow steps can allow for much quicker development and implementation of new steps. It gives you the ability to test for specific outputs when specific inputs are put into place. If you refactor the code or add in new steps you can run these tests to make sure that you haven't broken anything else in the code.

Finally, the workflow would all be moved to the new 
workflow interface that supports collaborative notebook interaction between all CHASE-CI users and developers. The CONNECT workflow would be presented as a series of steps in the UI where each step could easily be worked on. The workflow steps would be centralized in one location where every one working on the project could see them. It would allow for easier project management and the ability to scale out development. 

\section{Related Work} \label{sec:rw}

In addition to using multi-threaded software to leverage multiple GPUs for training neural networks, researchers can utilize their access to graphics cards for hardware-accelerated data visualization. In January 2019, Calit2 visualization researchers Joel Polizzi and Isaac Nealey used the CHASE-CI infrastructure to schedule and debug a scalable OpenGL-based visualization application \cite{c14} across 11 remote GPU nodes. They were able to lead a Virtual Reality content demonstration at University of California, Merced from an immersive visualization space at University of California, San Diego \cite{c15}, driving graphical displays in Merced with input from a motion tracked wand in San Diego with unnoticeable latency. Kubernetes object labeling conventions enabled straightforward targeting of specific nodes, and the high-bandwidth optical network between PRP sites enabled rapid inter-node communication and read speeds to the visual data. It is notable that graphics and machine learning processes can cohabitate, as remote researchers have the ability to run GPU compute jobs on the same hardware which is being used locally for visualization.

JupyterHub is also an integral part of the CHASE-CI Kubernetes GPU cluster. This software allows for a web based environment to automatically be generated per user on demand. The Jupyter Notebook instance that is generated is attached to a GPU on the cluster and can be automatically taken advantage of through framework software such as TensorFlow.  This process allows for quick development of code without the hassle of setting up any code or configuration locally on your system. Building on top of this, JupyterLab provides a great way to interact with notebooks and the node that it is connected to. It allows for collaboration between groups that need to use the same code and run it on a GPU resource.

There are also many other machine learning workflows that are taking advantage of the CHASE-CI infrastructure. These workflows are split up into different namespaces\footnote{Machine Learning Namespaces in CHASE-CI: http://ucsd-prp.gitlab.io/nautilus/namespaces/} to separate resources. Examples of these other projects include CARL-UCI, trying to apply the mechanisms in neuromodulation to reinforcement learning and use the signals from neuromodulation to modulate the learning and acting of the reinforcement learning algorithms. The GPU specific software they use includes Cuda, Pytorch, OpenAI gym, Conda, and Tensorboard.  Another workflow running on the cluster includes ECEWCSNG from UCSD.  This project includes building deep learning algorithms that can effectively combine data from other autonomous systems for safety applications. The GPUs are highly utilized from this workflow and the software used includes conda, tensorflow, caffe, numpy, and opencv. 

\section{Conclusion} \label{sec:conc}
There is a need for new workflow approaches that can be coupled with advanced cyberinfrastructure for efficient development and execution. In this paper, we presented the CHASE-CI software and hardware ecosystem that enables development and execution of collaborative, measurable, scalable and portable machine learning workflows.  

CHASE-CI provides an ideal infrastructure to rapidly test and build ML applications with a workflow layer that can optimize execution through Kubernetes. By coupling a dynamic cyberinfrastructure and the workflow process, we provided a new step-by-step workflow development approach for machine learning applications that drastically reduces execution bottlenecks by constantly measuring, learning, and informing every aspect of a machine learning workflow.

The representative segmentation workflow shows how careful measurement and analysis of a step by step ML workflow activity can enable reduction of execution time significantly through dynamic resource configuration via Kubernetes. The experimental results and performance measurements were presented using the CHASE-CI dashboard visualizations in Grafana.

\section*{Acknowledgment}

This work was supported in part by NSF-1730158 for Cognitive Hardware and Software Ecosystem Community Infrastructure (CHASE-CI), NSF-1541349 for The Pacific Research Platform (PRP), and NSF-1826967 for The Pacific Research Platform (NRP). The content is solely the responsibility of the authors and does not necessarily represent the official views of the funding agencies. The authors would also like to thank Shweta Purawat and Alok Singh for their participation in the discussions leading to this paper.

\vspace{12pt}


\begin{thebibliography}{00}
\bibitem{c1} L. Smarr, ``CHASE-CI: A Distributed Big Data Machine Learning Platform, Opening Talk With Professor Ken Kreutz-Delgado'', Qualcomm Institute University of California, San Diego, 2018.
\bibitem{c2} J. Graham, ``Building the Pacific Research Platform: A Workshop Towards Deploying a Science-Driven Regional ‘Big Data Freeway’'', Calit2, San Diego Supercomputer Center, and CITRIS, San Diego, 2015.
\bibitem{c3} ``Science DMZ Architecture'', Fasterdata.es.net, 2019. [Online]. 
\bibitem{c4} ``Kubernetes Documentation'', Kubernetes.io, 2019. [Online]. 
\bibitem{c5} ``Docker Documentation'', Docker Documentation, 2019. [Online]. 
\bibitem{c6} ``Device Manager Proposal'', GitHub, 2018. [Online]. 
\bibitem{c7} ``CUDA Runtime API: CUDA Toolkit Documentation'', Docs.nvidia.com, 2018. [Online]. 
\bibitem{c8} ``Rook'', Rook.io, 2019. [Online]. 
\bibitem{c9} S. Weil, A. Leung, S. Brandt and C. Maltzahn, ``RADOS: A Scalable, Reliable Storage Service for Petabyte-scale Storage Clusters'', 2019.
\bibitem{c10} ``Architecture — Ceph Documentation'', Docs.ceph.com, 2019. [Online].
\bibitem{c11} ``Grafana Documentation'', Grafana Labs, 2019. [Online].
\bibitem{c12} ``Overview - Prometheus'', Prometheus.io, 2019. [Online].
\bibitem{c13} Jim Basney, Terry Fleury, and Jeff Gaynor, ``CILogon: A Federated X.509 Certification Authority for CyberInfrastructure Logon,'' Concurrency and Computation: Practice and Experience, Volume 26, Issue 13, pages 2225-2239, September 2014. [Online]
\bibitem{c14} J. Schulze, A. Prudhomme, P. Weber and T. DeFanti, ``CalVR: An Advanced Open Source Virtual Reality Software Framework'', University of California San Diego, Calit2, La Jolla, CA, 2019.
\bibitem{c15} ``Infrastructure - Immersive Visualization Lab Wiki'', Ivl.calit2.net, 2015. [Online].
\bibitem{c16}Bosilovich, M. G., Robertson, F. R., Chen, J. (2011). Global energy and water budgets in MERRA. Journal of Climate, 24(22), 5721–5739. [Online].
\bibitem{c17}Gelaro, R., McCarty, W., Suárez, M. J., Todling, R., Molod, A., Takacs, L., … Zhao, B. (2017). The Modern-Era Retrospective Analysis for Research and Applications, Version 2 (MERRA-2). Journal of Climate, 30(14), 5419–5454. [Online]
\bibitem{c18}Rienecker, M. M., Suarez, M. J., Gelaro, R., Todling, R., Bacmeister, J., Liu, E., … Woollen, J. (2011). MERRA: NASA’s Modern-Era Retrospective Analysis for Research and Applications. Journal of Climate, 24(14), 3624–3648. [Online]
\bibitem{c19}Domenico, Ben et al. Thematic Real-time Environmental Distributed Data Services (THREDDS): Incorporating Interactive Analysis Tools into NSDL. Journal of Digital Information, [S.l.], v. 2, n. 4, feb. 2006. ISSN 1368-7506. [Online]
\bibitem{j17} M. Januszewski, J. Kornfeld, P.H. Li, A. Pope, T. Blakely, L. Lindsey, J. Maitin-Shepard, M. Tyka, W. Denk, V. Jain, ``High-precision automated reconstruction of neurons with flood-filling networks'', Nature Methods, Vol. 15.8, pp. 605-610 2018. DOI: 10.1038/s41592-018-0049-4
\bibitem{c20}Sellars, S., P. Nguyen, W. Chu, X. Gao, K.‐l. Hsu, and S. Sorooshian (2013), Computational Earth science: Big data transformed into insight, EOS Trans. AGU, 94(32), 277–278.
\bibitem{c21}Sellars SL, Kawzenuk B, Nguyen P, Ralph FM, Sorooshian S (2017a) Genesis, pathways, and terminations of intense global water vapor transport in association with large-scale climate patterns. Geophys Res Lett 44:12465–12475
\bibitem{c22}Sellars SL, Nguyen P, Kawzenuk B (2017b) The CONNected object, or CONNECT algorithm applied to National Aeronautics and Space Administration (NASA) Modern-era retrospective analysis for research and applications, version 2 (MERRA V2)—integrated water vapor from 1980 to 2016. UC San Diego Library Digital Collections. [Online]
\bibitem{Kepler} I. Altintas, C. Berkley, E. Jaeger, M.B. Jones , B. Ludaescher, S. Mock, Kepler An Extensible System for Design and Execution of Scientific Workflows, Proceedings of the 16th International Conference on Scientific and Statistical Database Management (SSDBM 2004), p. 423-424, 2004.
\bibitem{Kepler2} B. Ludaescher, I. Altintas, C. Berkley, D. Higgins, E. Jaeger, M. Jones M., E.A. Lee, J. Tao, Y. Zhao, ``Scientific Workflow Management and the Kepler System,'' Concurrency Computation Practice: and Experience, Vol.18, pp. 1039–1065, 2006.

\end{thebibliography}
\end{document}